\documentclass[journal=jacsat,manuscript=article]{achemso}
\usepackage[version=3]{mhchem} 
\usepackage{color}

\newcommand{\FigureLength}{7.7cm}
\newcommand{\red}[1]{{\color{black} #1   }}

\author{Zhen Wah Tan}
\affiliation[Cornell University]
{Department of Physics, Cornell University, 109 Clark Hall, Ithaca, New York 14853-2501, U.S.A.}
\alsoaffiliation[Institute of High Performance Computing]
{Institute of High
Performance Computing, 1 Fusionopolis Way, \#16-16 Connexis, Singapore
138632, Singapore}
\author{Jian-Sheng Wang}
\affiliation[National University of Singapore]
{Center for Computational Science and Engineering, and
Department of Physics, National University of Singapore, Singapore
117542}
\author{Chee Kwan Gan}
\email{ganck@ihpc.a-star.edu.sg}
\affiliation[Institute of High Performance Computing]
{Institute of High
Performance Computing, 1 Fusionopolis Way, \#16-16 Connexis, Singapore
138632, Singapore}

\title{
First-principles study of heat transport properties of graphene nanoribbons
}

\usepackage{graphicx}
\usepackage{epstopdf}
\mathchardef\mhyphen="2D

\begin{document}

\begin{abstract}

We use density-functional theory and the
nonequilibrium Green's
function method, as well as phonon dispersion calculations
to study the thermal conductance of graphene
nanoribbons with armchair and zigzag edges, with and without 
hydrogen passivation. We find that low-frequency phonon bands of the 
zigzag ribbons are more dispersive than
those of the armchair ribbons, and that this difference accounts for the anisotropy
in the thermal conductance of graphene nanoribbons.
Comparing our results with data on
\red{large-area}
graphene, edge effects are shown to contribute to thermal conductance, enhance
the anisotropy in thermal conductance of graphene nanoribbons, and
increase thermal conductance per unit width.
The edges with and without hydrogen-passivation
modify the atomic structure and ultimately influence the phonon
thermal transport differently for the two ribbon types.

\newpage
\red{Keywords: Graphene nanoribbons, thermal transport, nonequilibrium Green's function, phonon dispersion}

\end{abstract}
\maketitle

\section{Introduction}

Graphene nanoribbons (GNR), or strips of 
planar graphene\cite{Novoselov04v306},
 exhibit
several novel  
properties such
as large magnetoresistance\cite{Kim08v3} and high electron
mobility\cite{Bolotin08v146} which prove to be useful in nanoscale electronics.
\red{Graphene is known to have extremely high thermal conductivity\cite{Balandin08v8} that
depends on the flake size\cite{Ghosh08v92} and number of atomic planes\cite{Ghosh10v9}. 
GNR are expected to retain good thermal properties.
This, together with the availability of simpler
fabrication techniques than carbon nanotubes\cite{Barone06v6}, has caused GNR to gain much
attention in recent years due to their potential applications in modern devices.}
Experimental and theoretical work have
concluded that GNR can be metallic or semiconducting\cite{Nakada96v54,Son06v97,Li08v319},
and the energy gap varies with ribbon
width\cite{Nakada96v54,Han07v98,Chen07v40} and edge
orientation\cite{Son06v97}. These properties make semiconducting GNR an attractive
alternative channel material\cite{Obradovic06v88} capable of
producing smaller devices than those achievable with silicon.
\red{Metallic
GNR can also be used as interconnects\cite{Shao08v92},
and having circuits made entirely of GNR
holds the possibility of reducing or eliminating contact resistance\cite{Naeemi07v28}.}
For these reasons, various methods for fabricating GNR have been
developed\cite{Li08v319,Yang08v130,Jiao09v458}, including
scanning tunneling microscope lithography that offers precise
control of the structure of ribbons\cite{Tapaszto08v3}. Recent
developments have also enabled the production of sub-10-nanometer
GNR\cite{Bai09v9}, and greater control over edge
geometries\cite{Cai10v466}. Prototype GNR
field-effect transistors have also been developed and
characterized\cite{Yan07v7,Wang08v100}.

While the electronic properties of GNR have been explored in depth,
thermal transport in GNR---an important consideration
for thermoelectric performance and thermal management---has also gained more
attention recently\cite{Guo09v95,Hu09v9,Ghosh09v11,Xu10v81}. Several
studies have been made on the phonon
dispersions\cite{Qian09v46,Gillen09v80,Vandescuren08v78,Yamada08v77}
of GNR, while more recent articles have focused on phonon-mediated thermal
transport\cite{Lan09v79} of GNR. For example, Sevin\c{c}li and Cuniberti have investigated the thermoelectric performance of edge-disordered
GNR\cite{Sevincli10v81}. Hu {\it et
al.}\cite{Hu09v9} observed that ZGNR have higher thermal conductance
than AGNR of comparable widths, and attributed this to different
phonon scattering rates intrinsic to the ribbon geometry.
Several other studies have also confirmed the anisotropy in
thermal conductance of GNR\cite{Xu09v95,Bi10v150}, and a similar anisotropy
has been predicted by Jiang {\it et al.}\cite{Jiang09v79} in
graphene. In this Letter,
we employ density-functional theory (DFT) that accurately deals with the atomic and electronic structures,
the nonequilibrium Green function method that has been shown to be useful in studying the electron\cite{Brandbyge02v65} and heat
transport through nanostructures\cite{Wang06v74}, and also the standard theory of lattice dynamics to account for phonon transport through GNR.
We investigate how the atomic structures of GNR affect the phonon dispersions
and ultimately thermal transport, and explain the anisotropy
in thermal conductance of GNR.

\section{Methodology}
\label{sec:method}

\begin{figure}[htbp]
\centering\includegraphics[width=\FigureLength,clip]{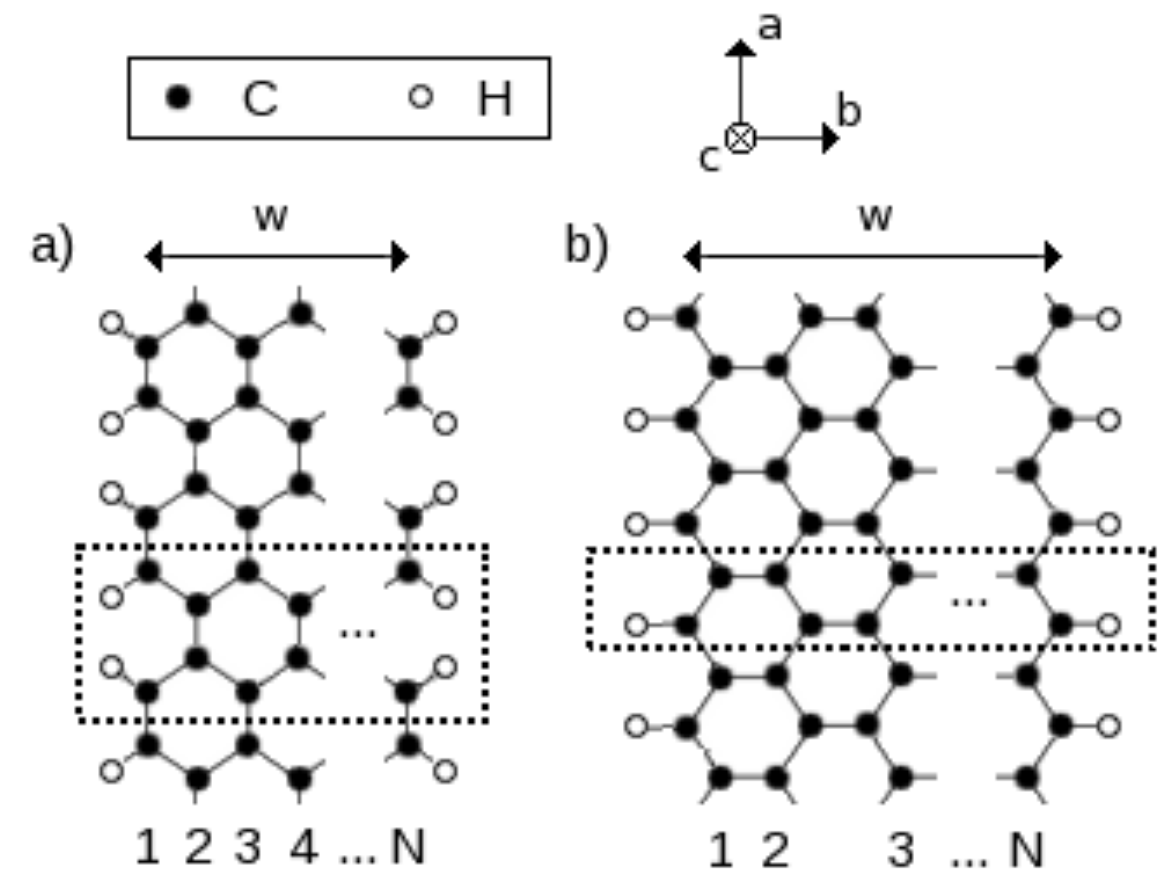}
\caption{Hydrogen-passivated graphene nanoribbons with (a)
armchair edges (AGNR-$N$), and (b) zigzag edges (ZGNR-$N$).  A
primitive unit cell is marked out in each case, and the ribbons are
periodic in the $a$-direction. The number $N$ denotes the number of C
atoms in the $b$-direction, and the ribbon width $W$ is given by the
maximal distance between C atoms in the $b$-direction.
}
\label{fig:GNR-definitions.eps}
\end{figure}

Because systems of sub-nanometer widths will become important with the
miniaturization of GNR-based devices, we study the thermal
conductance in narrow, pristine GNR by combining first-principles
density-functional calculations and the ballistic
nonequilibrium Green's function (NEGF)
method\cite{Wang06v74,Wang07v75}.
In this work, we investigate the thermal conductance of armchair GNR
(AGNR-$N$) of widths $N = 3, 4, 5, 6$
(refer to \ref{fig:GNR-definitions.eps}a for the meaning of $N$), as well as zigzag GNR
(ZGNR-$N$) of widths $N = 2, 3, 4, 5$
(\ref{fig:GNR-definitions.eps}b). We obtain the force-constant
matrix via DFT calculations implemented in SIESTA\cite{Soler02v14}.  We use
periodic boundary conditions on the orthorhombic supercell consisting
of nine primitive unit cells: The ribbons are periodic in the
$a$-direction (see Figure~\ref{fig:GNR-definitions.eps}), 
and each ribbon is separated from its nearest neighbors
by $16$ \AA\ of vacuum in the $b$- and $c$-directions. To facilitate 
NEGF calculations, we partition the supercell into three blocks with three primitive unit cells per block.
These three blocks are for the central junction, the left and right lead regions as described in Ref.~\cite{Wang07v75}.
We note that the same force constants required for NEGF calculations can be readily used to calculate
the phonon dispersion curves.

For DFT calculations, the local-density approximation with the exchange-correlation
functional due to Perdew and Zunger is used. We use a rather fine mesh cutoff
of $400$ Ry.
As demonstrated by Son {\it et al.}\cite{Son06v97}
and Gan {\it et al.}\cite{Gan10v81}, spin-polarization effects are
particularly important in ZGNR, therefore we use spin-polarized
calculations for ZGNR in this work. Nonspin-polarized calculations are used for
AGNR.

We first perform ionic relaxation of GNR using a conjugate-gradient
method. To obtain the best possible relaxed structure while ensuring
that the conjugate-gradient minimization converges, we use a force tolerance of
0.001 eV/\AA. To construct a force-constant matrix, we sequentially displace
each atom from its equilibrium position in
the $a$-, $b$- and $c$-directions by a distance of $0.015$~\AA, and
evaluate the forces acting on all atoms as a result of each
displacement\cite{Gan06v73}.  A central finite difference scheme
is used to evaluate the force-constant matrix, which is required by the 
NEGF and the phonon calculations.

Using the ballistic NEGF method,\cite{Wang06v74,Wang07v75} we then calculate the
phonon transmission coefficient $\tilde{T}$ for each GNR system.
Thermal conductance is evaluated using the Landauer formula:

\begin{equation}
\sigma(T) = \int_0^\infty \frac{d\omega}{2\pi} \ \hbar \omega \ \tilde{T}(\omega) \frac{\partial f}{\partial T},
\label{eq:Landauer-formula-dT}
\end{equation}
where the occupation distribution function $f(\omega, T) = 1/(e^{\hbar \omega / k_B T}-1)$.
It is important to note that \ref{eq:Landauer-formula-dT} implies that
at low temperatures, low-frequency modes are the dominant factor in thermal
transport since the derivative $\partial f/\partial T$ diminishes
rapidly with increasing $\omega$.

\newpage
\red{
We find the thermal conductance of GNR to be independent of ribbon length
in the ballistic regime, as noted previously by Lan {\it et
al.}\cite{Lan09v79}.
This is consistent with the theory of ballistic transport based on the Landauer formula, where diffusive behaviors
cannot be addressed. 
The ballistic assumption is valid for small systems, such as the ones studied in this work, where the
system size is much smaller than the graphene phonon mean free path of $\sim 775$~nm at room temperature\cite{Ghosh08v92}.
However, in the case of large-area graphene, diffusive (Umklapp-limited) 
scattering plays a significant role in reducing thermal conductivity,
as determined theoretically\cite{Nika09v79} and experimentally\cite{Freitag09v9,Nika09v94}.
}

\section{Results}
\label{sec:results}

\begin{figure}[htbp]
\centering\includegraphics[width=\FigureLength,clip]{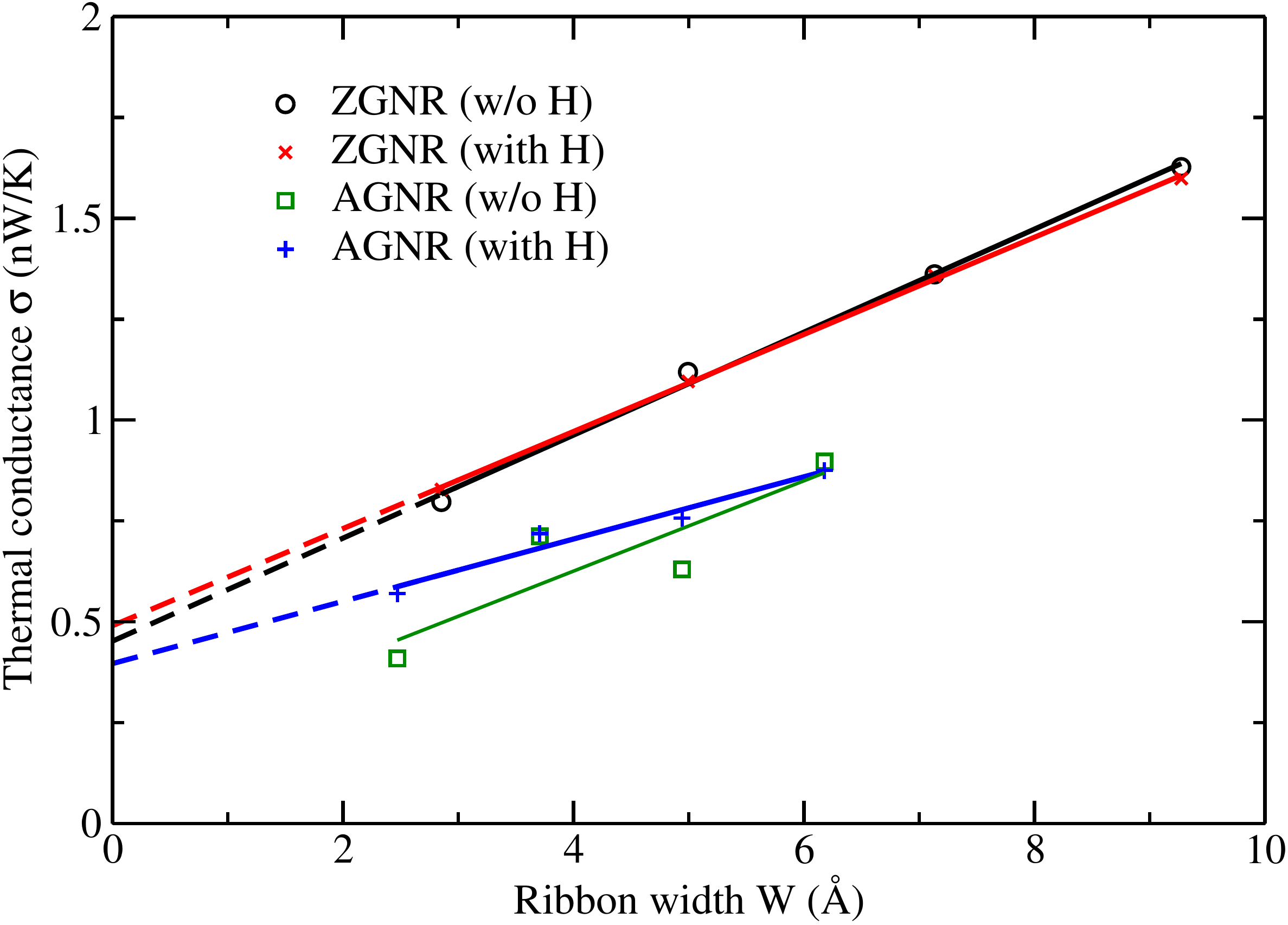}
\caption{(color online) Thermal conductance $\sigma$ (at 300~K) of GNR as a function of ribbon width $W$,
for ZGNR-$N$ with $N = 2,~3,~4,~5$ and AGNR-$N$ with $N = 3, 4, 5, 6$.
Data shown are for ZGNR without H-passivation, ZGNR with
H-passivation; AGNR without H-passivation, and AGNR with H-passivation---linear
fits have correlation coefficients 0.998, 1.000, 0.882 and 0.978, respectively.  Gradients of the
linear fits are 1.276, 1.203, 1.119 and 0.771~W/(m~K), while vertical
intercepts are 0.4523, 0.4906, 0.178 and 0.3967 nW/K, respectively.
}
\label{fig:thermalconductance-versus-width.eps}
\end{figure}

\begin{figure}[htbp]
\centering\includegraphics[width=\FigureLength,clip]{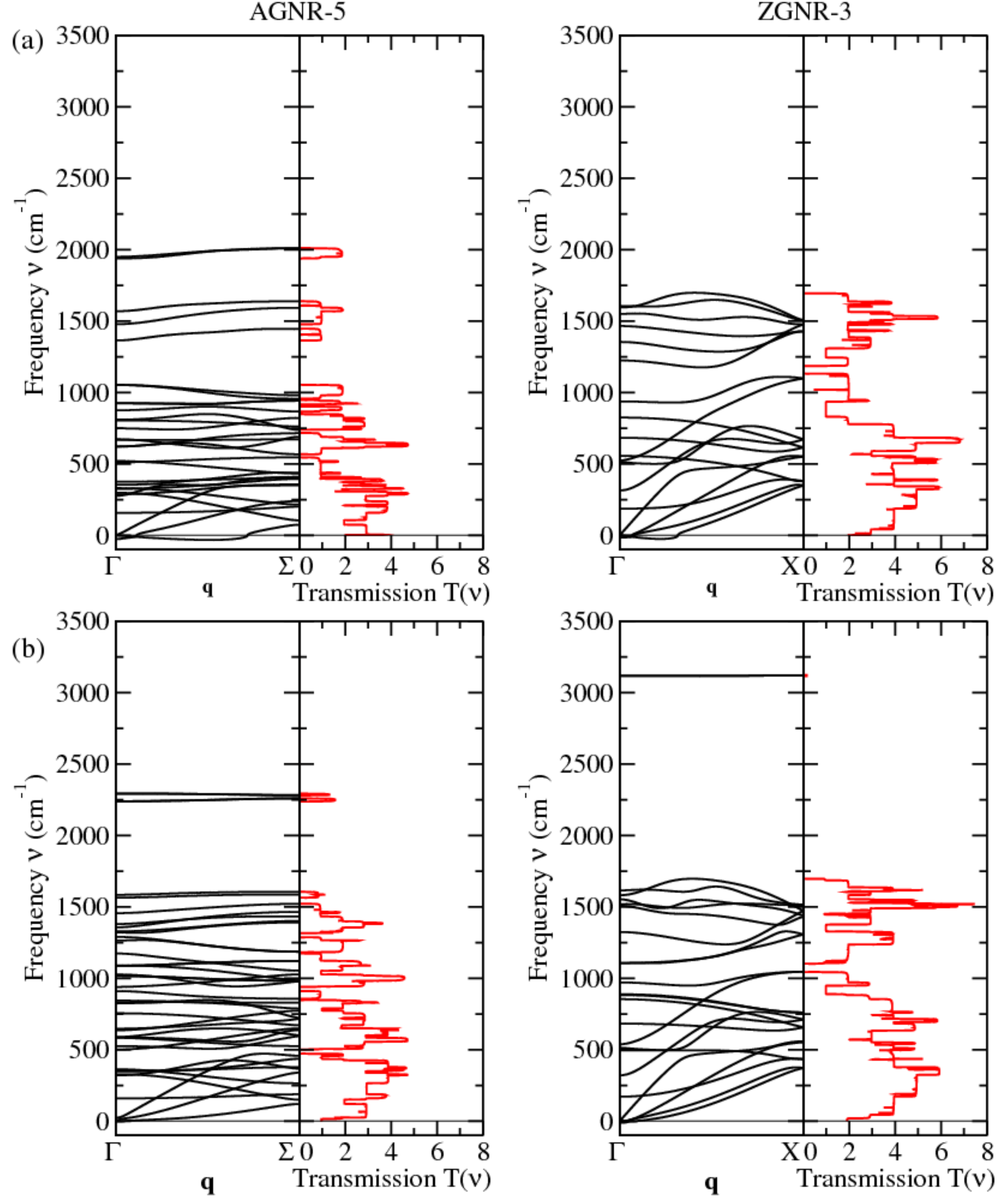}
\caption{(color online) Comparison between phonon
dispersions (left panel) and transmission coefficients (right panel) for
AGNR-5 and ZGNR-3, (a) without and (b) with H-passivation. Imaginary-frequency modes
are indicated by $\nu < 0$~cm$^{-1}$. The two
ribbons have comparable widths: $W_{\mathrm {AGNR\mhyphen5}} =
4.942$~\AA, $W_{\mathrm {ZGNR\mhyphen3}} = 4.993$~\AA.}
\label{fig:Sample-phonon-transm.eps}
\end{figure}

\ref{fig:thermalconductance-versus-width.eps} shows the thermal
conductance (at 300 K) of various GNR as a function of ribbon width
$W$.  In the limit of large width, the conductance of a ribbon is proportional
to its width, and our data exhibit a similar trend: For ZGNR with and
without H-passivation, and for AGNR with H-passivation, the thermal
conductance increases linearly with width, with high correlation
coefficients of $0.978$ and above. While thermal conductance of
AGNR without H-passivation generally increases with temperature, it does not
follow a strictly linear trend. On the other hand, on extrapolating
the linear trends to $W~=~0$~\AA, we see positive contributions to the
thermal conductance ($\sim\!0.5$~nW/K) that stem from edge effects.
This is consistent with the findings by Xu {\it et al.}\cite{Xu09v95},
which show a sharp increase in thermal conductance per unit width
as $W$ decreases from $1$~nm to $0.5$~nm, where edge
effects becomes more evident. \ref{fig:thermalconductance-versus-width.eps} also shows that
ZGNR generally have higher conductance than AGNR of comparable
widths. The main reason for this is due to the larger transmission coefficients
${\tilde T}(\nu)$ for the ZGNR compared to AGNR of comparable widths, as 
shown in \ref{fig:Sample-phonon-transm.eps}. This observation is
explained in the following paragraphs.

According to the NEGF formalism for 1D systems,\cite{Yamamoto04v92,Rego98v81,Wang07v98} the transmission
coefficient $\tilde{T}(\nu)$ is
the number of phonon modes at any
frequency $\nu = \omega/2\pi$. This can be readily
seen in \ref{fig:Sample-phonon-transm.eps}, where we compare the
phonon dispersions with the transmission coefficient for AGNR-5 and
ZGNR-3, with and without H-passivation. We see that $\tilde{T}(\nu)$
essentially equals to the number of bands at frequency $\nu$, 
except at very low frequencies. We expect that the discrepancy at low frequencies between $\tilde{T}(\nu)$ from NEGF calculations and 
that deduced from phonon bands is caused by intricate differences in the numerical treatments for these two methods. For example,
in our NEGF calculation, it is assumed that two atoms cease to interact with one another (i.e., the force constant between
them is set to zero) once they are are separated by
more than three primitive cells. However, we do not truncate force-constant matrix elements in the 
phonon calculations. The slightly imaginary modes near $\Gamma$ from the phonon calculations suggest that 
the GNR is unstable with respect to long wavelength periodic distortion, a fact that is consistent with the compressive
edge stresses that are inherent in the GNR edges\cite{Gan10v81}.
Despite small numerical uncertainties for extremely low frequency modes (which may be eliminated by
adopting a larger supercell at the expense of incurring larger computational costs), 
the data obtained for $\tilde{T}(\nu)$ is sufficiently accurate for conductance calculations. Since
$\tilde{T}(\nu) $ is equal to the number of phonon modes present, dispersive
phonon bands would generally lead to larger values of $\tilde{T}(\nu)$
than less dispersive bands.

The dispersion relations in \ref{fig:Sample-phonon-transm.eps} indicate
that low-frequency phonon bands in ZGNR-$3$ tend to be more dispersive
than those in AGNR-$5$.  The ribbons have comparable widths,
with $W_{\mathrm {ZGNR\mhyphen3}} = 4.993$~\AA, and
$W_{\mathrm{AGNR\mhyphen5}} = 4.942$~\AA. The dispersive bands in ZGNR-3 would
give rise to higher $\tilde{T}(\nu)$ in the low-frequency regime:
low-frequency modes play a dominant
role in thermal conductance at room temperature. Thus, the fact that
low-frequency bands in ZGNR-3 are more dispersive than those in AGNR-5
explains the anisotropy in thermal conductance, even in the ballistic limit.

\begin{figure}[htbp]
\centering\includegraphics[width=\FigureLength,clip]{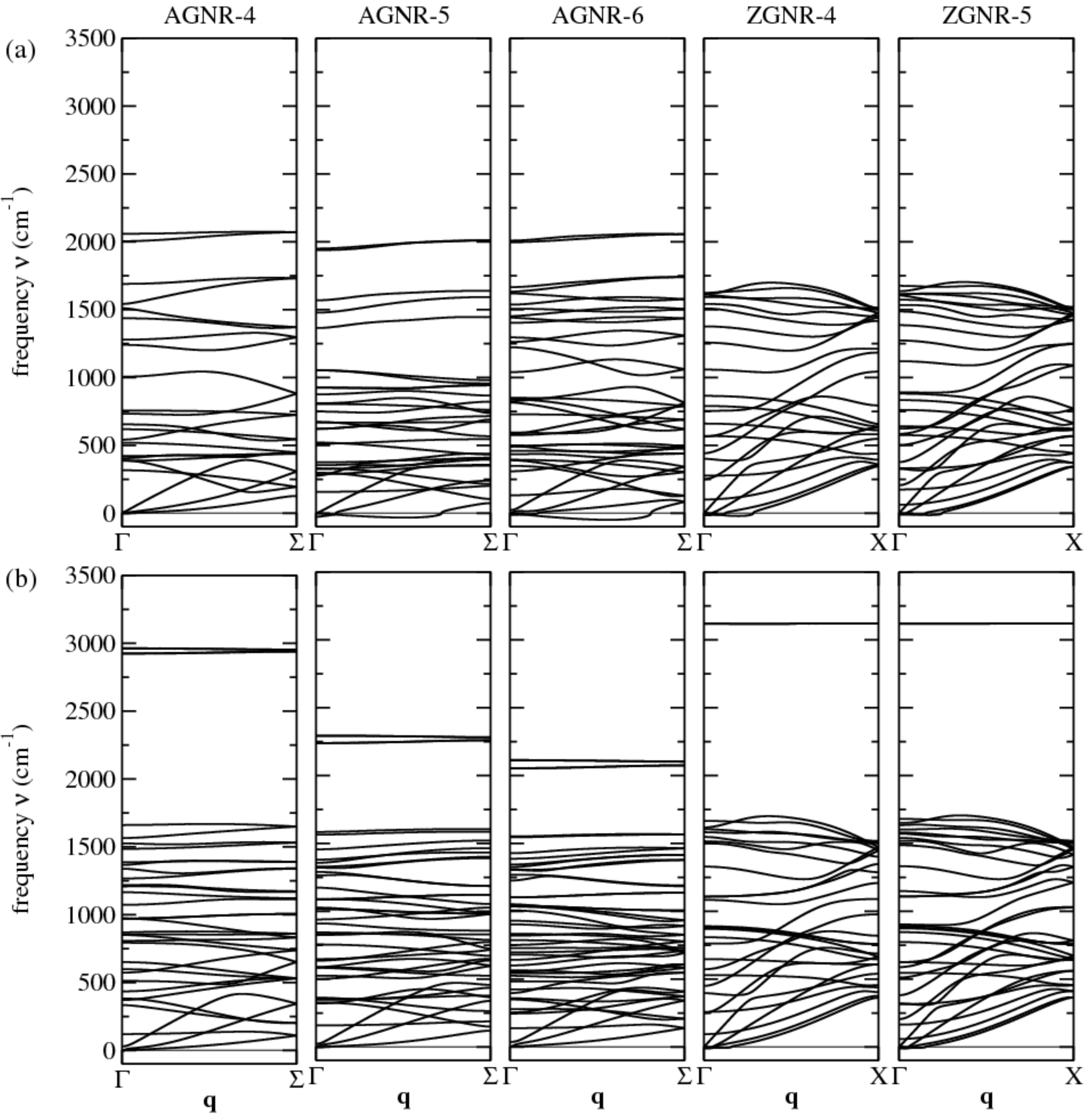}
\caption{Phonon dispersion plots for AGNR-4,~5,~6 and ZGNR-4,~5 (a) without
H-passivation and (b) with H-passivation. Imaginary frequencies are indicated by $\nu < 0$~cm$^{-1}$.}
\label{fig:All-phonons.eps}
\end{figure}

Since the transmission coefficients are intimately related to the
phonon dispersions, we have systematically studied the phonon dispersion
of the ribbons of different width and orientations (i.e.,
AGNR-$N$ and ZGNR-$N$), with and without
hydrogen passivation, where the results are summarized in
\ref{fig:All-phonons.eps}. The
dispersion trends, cut-off frequencies, and C-H stretching mode
frequencies agree very well with previously published data calculated
using REBOII and MO/8
simulations\cite{Vandescuren08v78,Yamada08v77}. Also, DFT calculations
by Gillen {\it et al.}\cite{Gillen09v80} yielded very similar phonon
dispersion trends for ZGNR, with cut-off frequencies of
$\sim\!\!1650$~cm$^{-1}$, and C--H stretching mode frequencies of
$\sim\!3100$~cm$^{-1}$, while we obtain $\sim\!\!1700$~cm$^{-1}$ and
$\sim\!3120$~cm$^{-1}$ respectively.

It is interesting to note that in most of the test cases, the lowest phonon
modes are found to soften and have imaginary frequencies---an indication of
instability---near $\Gamma$. These imaginary-frequency modes (denoted by $\nu < 0$~cm$^{-1}$) 
are shown in
\ref{fig:All-phonons.eps}. The lowest modes are edge
waves with out-of-plane displacements, and the unstable edge rippling
generally occurs over a range of about 10\% the Brillouin zone size,
corresponding to wavelengths larger than 2~nm (4~nm) for ZGNR (AGNR). This result is consistent with
a study by Shenoy~{\it et al.}\cite{Shenoy08v101}, which indicates that GNR
form edge ripples with wavelengths of about 8~nm due to edge stresses. 

The phonon plots in \ref{fig:All-phonons.eps} show that in general, ZGNR
have low-frequency bands that are more dispersive than those in AGNR.
The reasoning for ZGNR-3 having higher thermal conductance than AGNR-5 can thus be 
applied also to relate the anisotropic behavior of GNR: More dispersive low-frequency bands in
ZGNR result in larger transmission coefficients $\tilde{T}(\nu)$ as compared to AGNR in
the low-frequency regime, thereby giving rise to higher thermal conductance in ZGNR.

A similar anisotropy in thermal conductance in graphene sheets has
been pointed out by Jiang {\it et al.}\cite{Jiang09v79}, using the
valence force field model. The study found a thermal conductance
per unit width, $\sigma/W$ = 0.348 and 0.352~W/(m~K)
in the armchair and zigzag directions, respectively.
Similar findings have been obtained by Hu {\it et
al.}\cite{Hu09v9} and Guo {\it et al.}\cite{Guo09v95} through
molecular dynamics simulations for GNR with larger $N$.  Hu {\it
et al.} attributed the difference in thermal conductance between AGNR
and ZGNR to the presence of different phonon scattering mechanisms.
However, since we observe that the anisotropy persists in the ballistic regime,
we suggest that the anisotropy in thermal conductance 
could also be due to the anisotropy in
phonon dispersion for GNR.

On the other hand, the fitted gradients of our $\sigma$ versus $W$ plots for
H-passivated ribbons are 0.771 and 1.203~W/(m~K), respectively. This
suggests that edge effects increase the $\sigma/W$ ratio, and magnify the
anisotropy in thermal conductance for GNR. These trends indicate that
arrays of narrow ZGNR may be most efficient in thermal management,
whereas wide AGNR may optimize thermoelectric performance
of graphene-based transistors, with its lower thermal conductance.

\begin{table}[h]
\begin{center}
\begin{tabular}{c|c|c|c|c|c|c}
\hline\hline
$ N $ & \multicolumn{3}{c|}{ AGNR }{}& \multicolumn{3}{c}{ ZGNR }{}  \\
\hline
& $\sigma_{\mathrm {w/o H}}$ & $\sigma_{\mathrm {with H}}$ & $\Delta$ & $\sigma_{\mathrm {w/o H}}$ & $\sigma_{\mathrm {with H}}$ & $\Delta$ \\
\hline
$\ 2 \ $ &  &  &  & \ 0.797 \  & \ 0.828 \ & \ 3.9\% \   \\
$\ 3 \ $ & \ 0.409 \  & \ 0.570 \ & \ 39.4\% \  & \ 1.119 \  & \ 1.096\ & \ -2.1\% \  \\
$\ 4 \ $ & \ 0.712 \  & \ 0.719 \ & \ 1\% \  & \ 1.361 \  & \ 1.358 \ & \ -0.2\% \  \\
$\ 5 \ $ & \ 0.630 \  & \ 0.757\ & \ 20.1\% \  & \ 1.627\  & \ 1.599\ & \ -1.7\% \  \\
$\ 6 \ $ & \ 0.897 \  & \ 0.875 \ & \ -2.5\% \  &  &  &  \\
\hline
\end{tabular}
\end{center}
\caption{Thermal conductance data of GNR. Thermal conductance at 300~K is $\sigma$, in units of nW/K. The conductance
shift due to H-passivation is $\Delta = (\sigma_{\mathrm{with\,H}} - \sigma_{\mathrm{w/o\,H}}) / \sigma_{\mathrm{w/o\,H}}$.}
\label{table:thermal-conductance-at-300K}
\end{table}

In considering the effects of H-passivation on $\sigma$, we define the
thermal conductance shift by
$\Delta=(\sigma_{\mathrm{with\,H}}-\sigma_{\mathrm{w/o\,H}})/\sigma_{\mathrm{w/o\,H}}$.
Table \ref{table:thermal-conductance-at-300K} shows that while
$\Delta$ stays below $4\%$ for ZGNR, $\Delta$ can reach as high as
$39\%$ for AGNR.  This disparity can be traced to the modification of
phonon modes with H-passivation.

\ref{fig:All-phonons.eps} shows
that H-passivation induces little change to the dispersion relations
for ZGNR, compared to AGNR.  For ZGNR, despite the addition of six
phonon bands (due to two extra H atoms), these bands have high
frequencies ($\nu > 700$~cm$^{-1}$) and thus do not contribute
significantly to thermal conductance at room temperature. On the other
hand, for AGNR, we see that H-passivation causes more drastic changes to phonon
dispersions. Besides having twelve more phonon bands due to
the four H atoms, the phonon bands also generally shift towards lower
frequencies.  To explain the latter, we note that H-passivation
reduces the strong C--C triple bonds at the armrests of
AGNR\cite{Kawai00v62,Koskinen08v101} to a much weaker edge
bond. However, in ZGNR, only one in two edge C atoms have dangling bonds, thus edge C--C
bonds are weaker than double bonds, and relax only slightly upon
H-passivation. Bond relaxation causes a reduction in force constants
and phonon frequencies, and this change takes place more significantly
in AGNR than in ZGNR.

It should also be noted that for AGNR with and without H-passivation, the
thermal conductance varies non-trivially with ribbon width, with
larger variations observed in the cases without H-passivation. Previous
studies on AGNR have shown that the variation of energy
gaps\cite{Fujita97v66,Kawai00v62,Son06v97} and edge
energies\cite{Gan10v81} with ribbon width can be grouped into three
distinct families (with respect to $N$ mod 3). However, we have been
unable to verify such trends in thermal conductance for AGNR due to
huge computational costs for systems of larger $N$.

\section{Conclusions}

We have studied phonon-mediated thermal conductance of graphene
nanoribbons with armchair (AGNR) and zigzag (ZGNR) edges of
sub-nanometer ribbon widths through density-functional
calculations, a nonequilibrium Green's function method, and phonon calculations.  ZGNR was
found to have higher thermal conductance than AGNR of comparable
widths, due to an anisotropy in the phonon dispersion for GNR:
low-frequency bands in ZGNR are more dispersive than those in AGNR.
Edge effects were found to produce a positive contribution
($\sim\!0.5$~nW/K) to thermal conductance, increase
thermal conductance per unit width, and cause ZGNR to have a
significantly higher thermal conductance per unit width
($\sim\!\!1.2$~W/(m~K)) than AGNR ($\sim\!0.77$~W/(m~K)). These facts
suggest that narrow ZGNR can act as good thermal conductors for thermal
management, while wide AGNR may be a better candidate for application in
nanoscale devices.
Thermal conductance of AGNR and ZGNR change differently with
hydrogen-passivation at the edges: Edge C--C bonds in AGNR are relaxed
by a greater extent than those in ZGNR, thus phonon frequencies in
AGNR are more significantly reduced than those in ZGNR.  It remains
interesting to determine if thermal conductance in AGNR follows the
same trends as that for energy gap and edge energy, as this will yield
further insight on thermal transport mechanisms in AGNR. This may shed some light
in understanding the thermoelectric behavior of AGNR, and
ultimately be useful for applications in nanoscale electronics.

Finally, we note that the combination of density-functional theory,
nonequilibrium Green's function method and phonon dispersion calculations
reveal, for the first time, the peculiar heat transport properties that could be traced to
the intricate interplay between the atomic structures, the transmission
coefficients and the dispersiveness of the phonon bands. Our study provided
a natural link between these quantities and explained
the anisotropy observed in the thermal conductance of zigzag and armchair GNR.

\section{Acknowledgments}

The authors gratefully acknowledge invaluable discussions with Julian D.
Gale, Jinghua Lan, Gang Wu, and Jinwu Jiang.
\providecommand*\mcitethebibliography{\thebibliography}
\csname @ifundefined\endcsname{endmcitethebibliography}
  {\let\endmcitethebibliography\endthebibliography}{}

\end{document}